\documentclass[11pt,twoside]{article}
\usepackage{CAGN2019}
\usepackage{graphicx}

\usepackage[T1]{fontenc} 

\usepackage{latexsym}
\usepackage{verbatim}
\usepackage{ifpdf} 
\usepackage{color} 
\usepackage{ulem}

\newcommand{\temp}{$T_{\rm e}$}
\newcommand\ion[2]{#1$\;${\scshape{#2}}}

\ifpdf  
      \DeclareGraphicsExtensions{.pdf,.png,.jpg}  
\else  
      \DeclareGraphicsExtensions{.eps}  
\fi 

\begin{document}

\vskip 1.0cm
\markboth{G.~Delgado-Inglada, A. Medina-Amayo, \& G. Stasi\'nska}{Ionization correction factors for ionized nebulae}
\pagestyle{myheadings}

\vspace*{0.5cm}
\parindent 0pt{Invited Review}


\vspace*{0.5cm}
\title{Ionization correction factors for ionized nebulae}

\author{G.~Delgado-Inglada$^1$, A.~Medina-Amayo$^1$, and G.~Stasi\'nska$^2$}
\affil{$^1$Instituto de Astronom\'ia, Universidad Nacional Aut\'onoma de M\'exico, Mexico (email: \texttt{gdelgado@astro.unam.mx)}\\
$^2$LUTH, Observatoire de Paris, CNRS, UniversitÈ Paris Diderot; Place Jules Janssen, F-92190 Meudon, France}

\begin{abstract}
In this paper we discuss the calculation of chemical abundances in planetary nebulae and \ion{H}{ii} regions through ionization correction factors (ICFs). We review the first ICFs proposed in the literature based on ionization potential similarities and we present the most recent ICFs derived from large sample of photoionization models. We also discuss some of the considerations that have to be kept in mind when using ICFs to compute the chemical composition of ionized nebulae.


\end{abstract}

\section{Introduction}
The chemical abundances of planetary nebulae (PNe) and \ion{H}{ii} regions may be computed using different approaches \citep[see, e.g.,][]{sta2002,pei2017}. The so-called direct method consists of adding up all the ionic abundances of the ions present in the nebulae:
\begin{equation}
\frac{{\rm X}}{{\rm H}} = \frac{{\rm X}^{+}}{{\rm H}^{+}} + \frac{{\rm X}^{++}}{{\rm H}^{+}} + \frac{{\rm X}^{+3}}{{\rm H}^{+}} +...,
\label{icfs1}
\end{equation}
and it requires to know the physical conditions: the electron temperature and density ($T_{\rm e}$ and $n_{\rm e}$, respectively) in the regions where the different lines are emitted. When one cannot observe all the ions present in a nebula (either because the emission lines are too weak or because the observed wavelength range does not include all the involved lines), a correction must be made using ionization correction factors (ICFs):
\begin{equation}
\frac{{\rm X}}{{\rm H}} = \Sigma_{\rm obs.}\frac{{\rm X}^{+i}}{{\rm H}^{+}} \times {\rm ICF},
\label{icfs1}
\end{equation}

where $\Sigma_{\rm obs.}$ represents the sum of all the ionic abundances that can be computed from observations. Therefore, ICFs account for the contribution of unobserved ions and they may be obtained from ionization potential similarities or from photoionization models (see Sections~\ref{icfs_IP} and \ref{icfs_models}).

\section{ICFs based on ionization potential similarities}
\label{icfs_IP}
The first ICFs were proposed 50 years ago. In the paper we are celebrating in this meeting, \citep{pei1969}, the authors computed the total abundances of various elements in Orion, M8, and M17 based on ionization potential similarities. Three of the expressions proposed by these authors that are still commonly used to derive nitrogen, neon, and sulfur abundances are:
\begin{equation}
{\rm N}/{\rm O} = {\rm N}^{+}/{\rm O}^{+},
\label{icfs1}
\end{equation}
\begin{equation}
{\rm Ne}/{\rm O} = {\rm Ne}^{++}/{\rm O}^{++},
\label{icfs2}
\end{equation}
and
\begin{equation}
{\rm S}/{\rm O} = ({\rm S}^{+}+{\rm S}^{++})/{\rm O}^{+}.
\label{icfs3}
\end{equation}
The ionization potentials of N$^{+}$, Ne$^{++}$, S$^{++}$ are 29.6, 63.4, and 34.8 eV, and those of O$^{+}$ and O$^{++}$ are 35.1 eV and 54.9 eV, respectively. The fact that N$^{+}$ and O$^{+}$, Ne$^{++}$ and O$^{++}$, and S$^{++}$ and O$^{+}$ have similar ionization potentials has been used to propose these simple ICFs. 

\citet{pei1969} argued that the correction scheme was correct because the derived abundances in the three regions of Orion nebula studied by them were similar. But the observational basis was scarce at that time. Using recent photoionization models one finds that these simple expressions are not always valid and that new ICFs are needed to obtain more reliable abundances. 

\begin{figure}[!t]
\begin{center}
\includegraphics[height=5cm, trim = {-45 5 35 5}, clip]{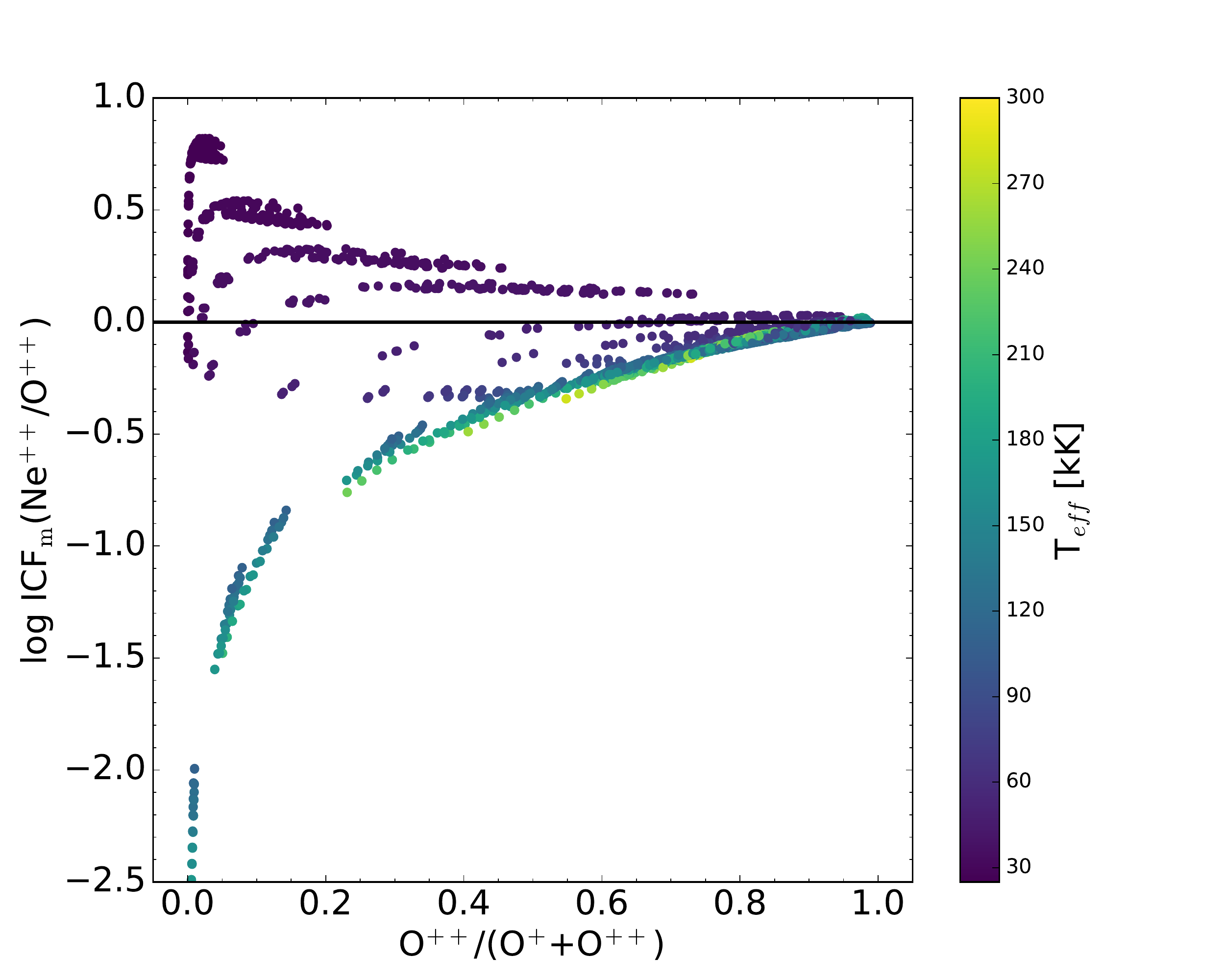}
\includegraphics[height=5cm, trim = {-45 5 35 5}, clip]{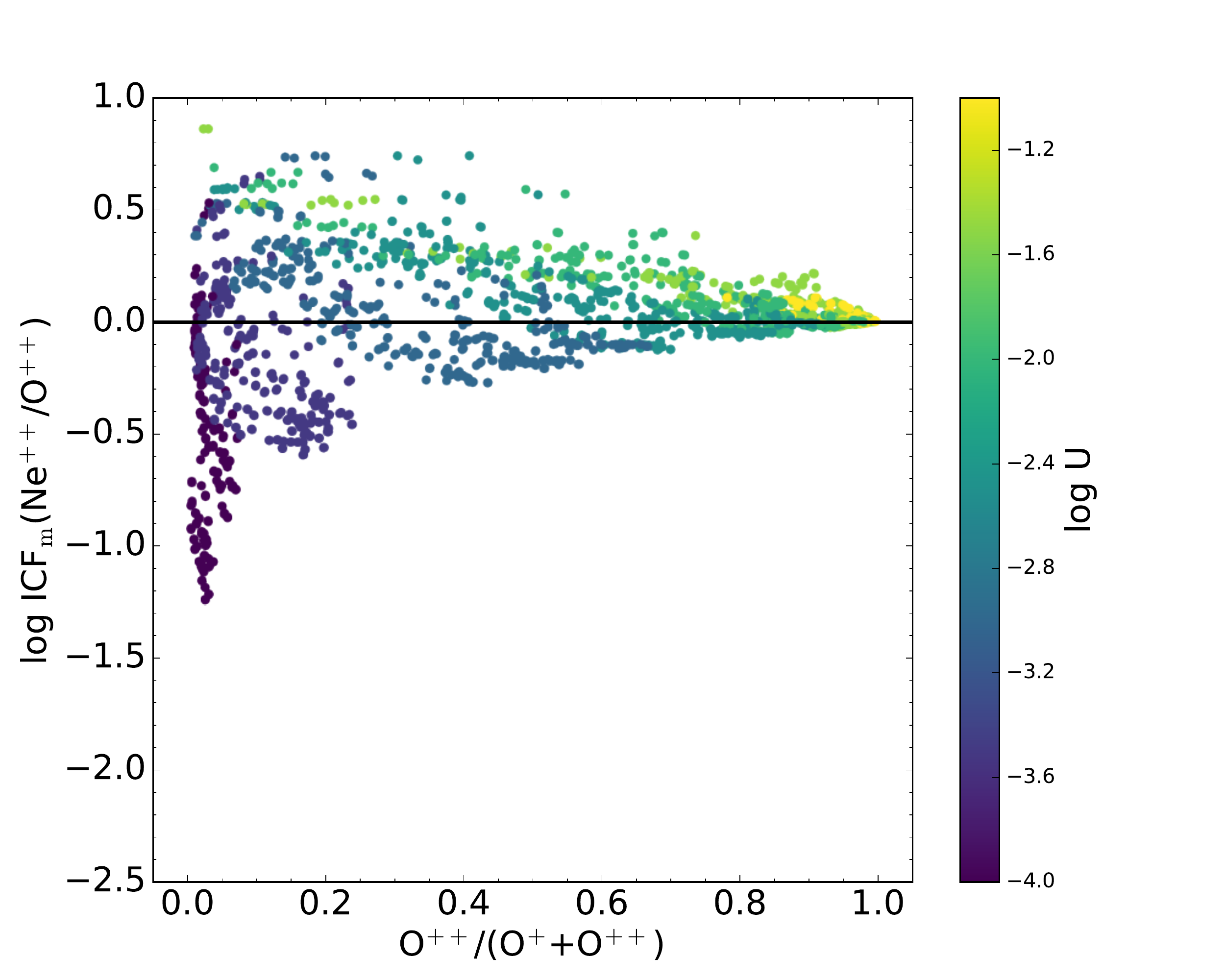}
\includegraphics[height=5cm, trim = {-45 5 35 5}, clip]{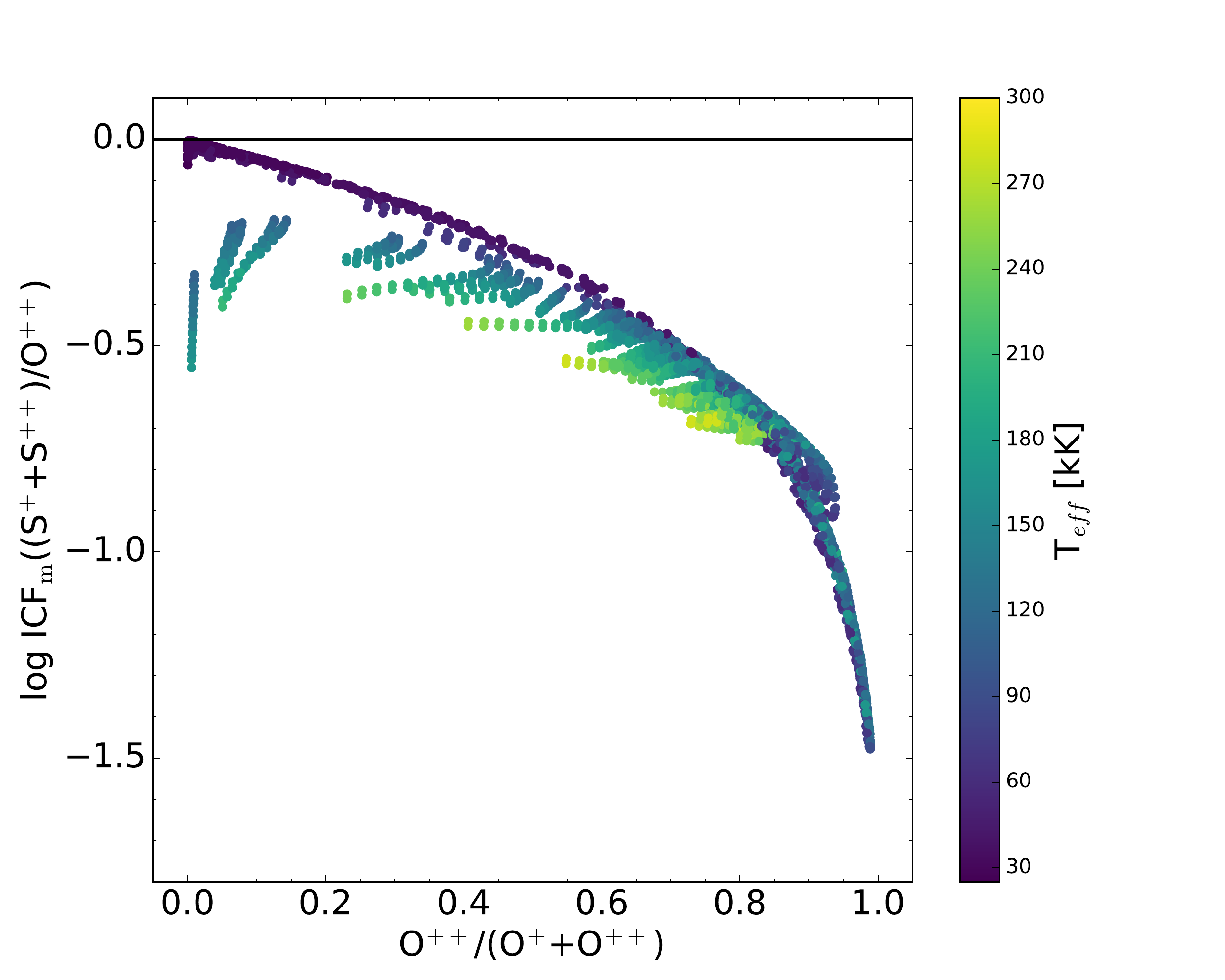}
\includegraphics[height=5cm, trim = {-45 5 35 5}, clip]{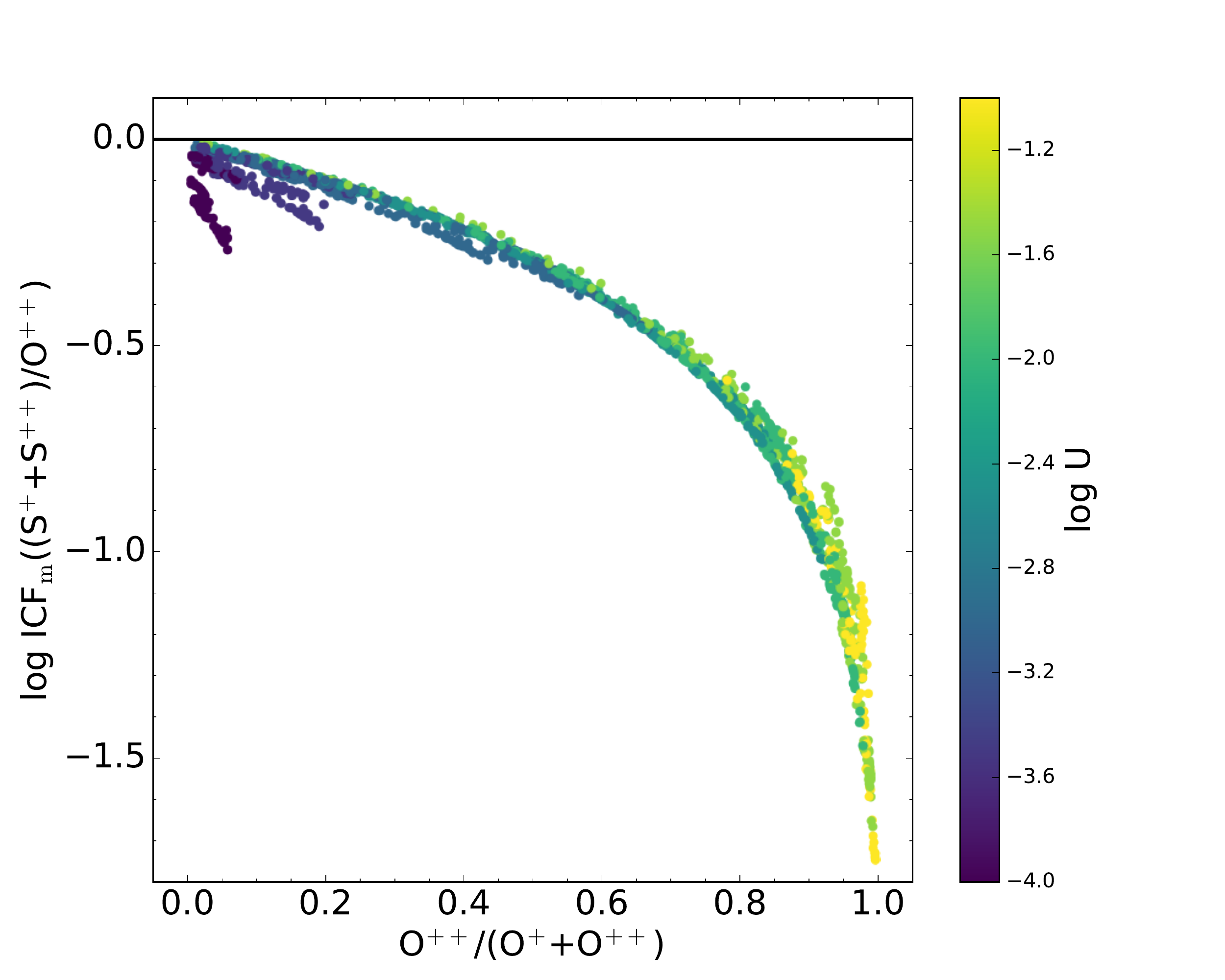}
\caption{Values of $x$(O$^{++}$)/$x$(Ne$^{++}$) (upper panels) and $x$(O$^{+}$)/($x$(S$^{+}$)+$x$(S$^{++}$)) (lower panels) as a function of O$^{++}$/(O$^{+}$+O$^{++}$) for our sample of photoionization models representative of PNe (left panels) and extragalactic \ion{H}{ii} regions (right panels). The colorbars run from low lo high values of the effective temperature of the central star (for the PN models) and of the ionization parameter (for the extragalactic \ion{H}{ii} regions).}
\label{fig1}
\end{center}
\end{figure}

Figure~\ref{fig1} illustrates this for neon and sulfur. It is obvious that the ICFs based on ionization potential similarities (represented by the solid line) lead to incorrect values of Ne/O and S/O in most of the cases. In particular, Equation~\ref{icfs3} systematically overestimates S/O values whereas Equation~\ref{icfs2} only provide reliable values of Ne/O for objects with O$^{++}$/(O$^{+}$+O$^{++}$) $\geq$ 0.6.

\section{ICFs based on photoionization models}
\label{icfs_models}

The ICFs derived from photoionization models are, in principle, more reliable because photoionization codes include the physics involved in ionized nebulae (proper photoionization and recombination rates, charge transfer reactions) that may change the distributions of ions in nebulae. However, one must remember that the atomic physics included in photoionization codes can be incorrect and also that some of the assumptions made to construct the models may be too simplistic (for example, the real structure of nebulae may be different than the one assumed). ICFs from photoionization models have been derived by \citet{kin1994, del2014} for planetary nebulae and by \citet{mat1985, sta1978, mat1991, gru1992, izo2006, dor2016} for \ion{H}{ii} regions. 

A few years ago we started a project to obtain new ICFs for \ion{H}{ii} regions and PNe from large grids of photoionization models computed with CLOUDY \citep{fer2013}. The database generated is available at the web page {\tt https://sites.google.com/site/mexicanmillionmodels/} and it was presented and described with detailed by \citet{mor2015}.

Our approach is to compute analytical expressions for the ICFs that depend on O$^{++}$/(O$^{+}$+O$^{++}$) (and He$^{++}$/(He$^{+}$+He$^{++}$) for planetary nebulae). We have chosen these abundance ratios because they can be easily computed from strong emission lines. In total we computed ICFs for He, C, N, O, Ne, S, Cl, Ar, Ni, Na, K, and Ca to be used in PNe (\citealt{del2014, del2016}, Medina-Amayo et al. 2019a, in prep.) and for C, N, O, Ne, S, Cl, Ar to be used in extragalactic \ion{H}{ii} regions (Medina-Amayo et al. 2019b, in prep.). 

\section{ICFs for PNe}
\label{icfs_pne}

\citet{del2014} described the grid of photoionization models that has been used to derive new ICFs in PNe. The input parameters of the models cover a wide range of values so that the grid is representative of many observed PNe. For example, the grid covers from 25000 to 300000 K in effective temperature, from $3\times10^{15}$ to $3\times10^{18}$ cm in inner radius, from 30 to 300000 cm$^{-3}$ in hydrogen density, and from 200 to 17800 L$_\odot$ in stellar luminosity. We have also checked that changes in e.g., the chemical composition of the nebula, the density law of the gas, or the presence or absence of dust did not affect the ICFs derived. 

One strength of the ICFs derived by \citet{del2014} is that together with the analytical expressions of the ICFs we provide analytical expressions for the uncertainties associated with each ICF. In general these uncertainties are not taken into account and they may be not negligible. For example, the uncertainty in $\log$(O/H), $\log$(N/O), and $\log$(Ar/H) associated to the ICF are $\sim$0.1, $\sim$0.2, and $\sim^{+0.18}_{-0.30}$ dex, respectively.

However, it must be noted that the grid of photoionization models considered was probably too extended for commonly observed planetary nebulae and the procedure to derive the uncertainties can be  improved. A future study will revise the work by \citet{del2014}.

\section{ICFs for extragalactic \ion{H}{ii} regions}
\label{icfs_hii}

\begin{figure}
\begin{center}
\includegraphics[height=6.5cm]{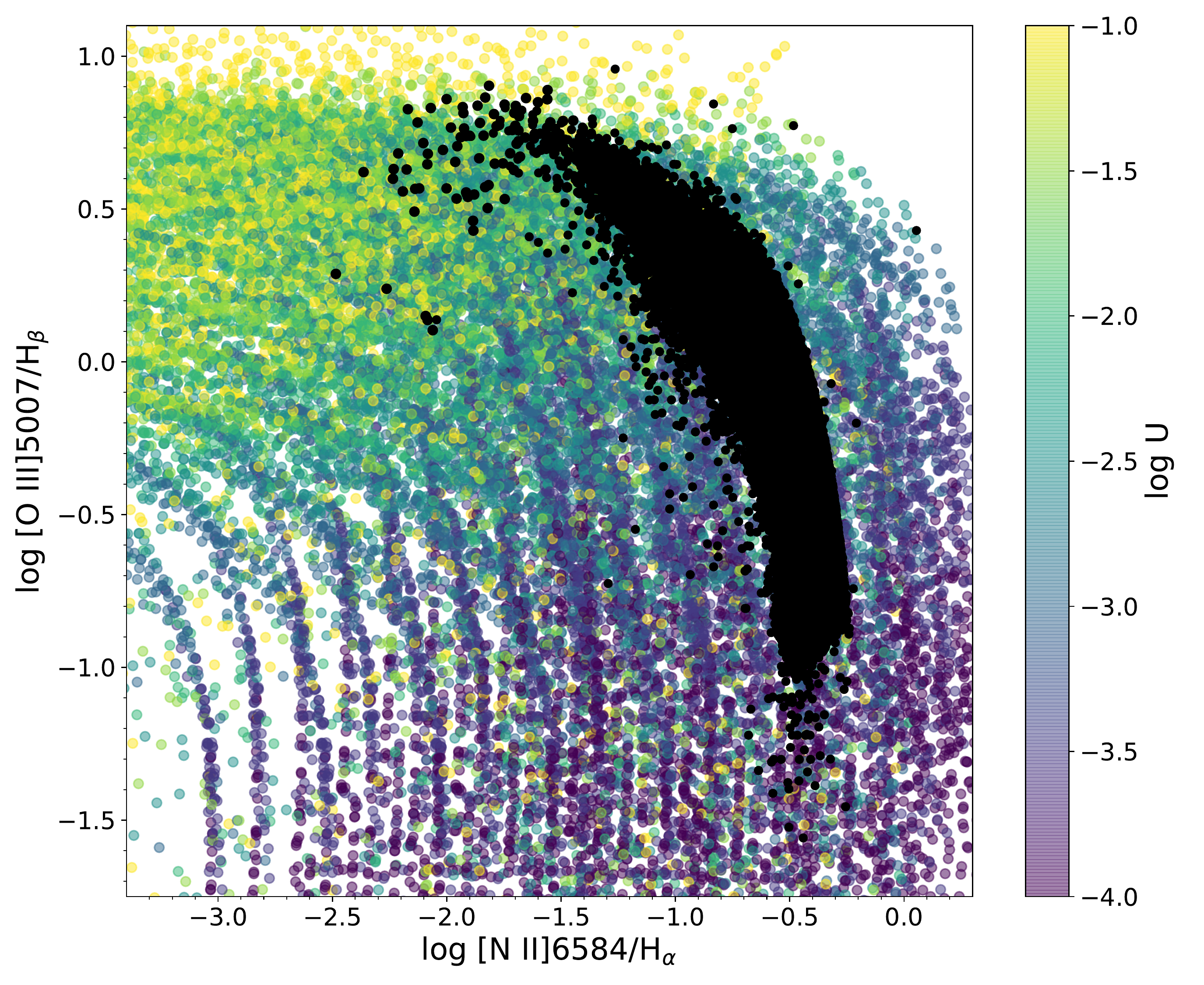}
\caption{Values of [\ion{O}{iii}] $\lambda$5007/H$\beta$ as a function of [\ion{N}{ii}] $\lambda$6584/H$\beta$ (the BPT diagram). The color dots represent the initial sample of 31500 models from \citet{val2016} whereas the black dots represent the observational sample of extragalactic \ion{H}{ii} regions (see the text for more details).} 
\label{fig2}
\end{center}
\end{figure}

Medina-Amayo et al. (2019, in prep.) present new ICFs for C, N, Ne, S, Cl, and Ar adequate to compute chemical abundances in extragalactic \ion{H}{ii} regions. The sample of models was selected from the grid of photoionization models computed by \citet{val2016} for giant \ion{H}{ii} regions. A large set of observations of extragalactic \ion{H}{ii} regions were used to constrain the sample of models from which the ICFs have been derived.  It includes blue compact galaxies, giant \ion{H}{ii} regions, and galaxies from the Sloan Digital Sky Survey \citep{str2002}. The data were taken from the same references used by \citet{val2016} and also from \citet{ber2013}, \citet{bre2011a} and \citet{bre2011b}. According to those observations, the sample was restricted in the N/O vs. O/H plane, the ionization parameter vs. O/H plane and the [\ion{O}{iii}] $\lambda$5007/H$\beta$ vs. [\ion{N}{ii}] $\lambda$6584/H$\beta$ plane. Fig.~\ref{fig2} shows the BPT diagram ([\ion{O}{iii}] $\lambda$5007/H$\beta$ vs [\ion{N}{ii}] $\lambda$6584/H$\beta$) for the sample of models from \citet[][color dots, 31500 models]{val2016} and for the sample of observations used by Medina-Amayo et al. (black dots, $\sim$133500 objects).

\begin{figure}
\begin{center}
\includegraphics[height=6.5cm]{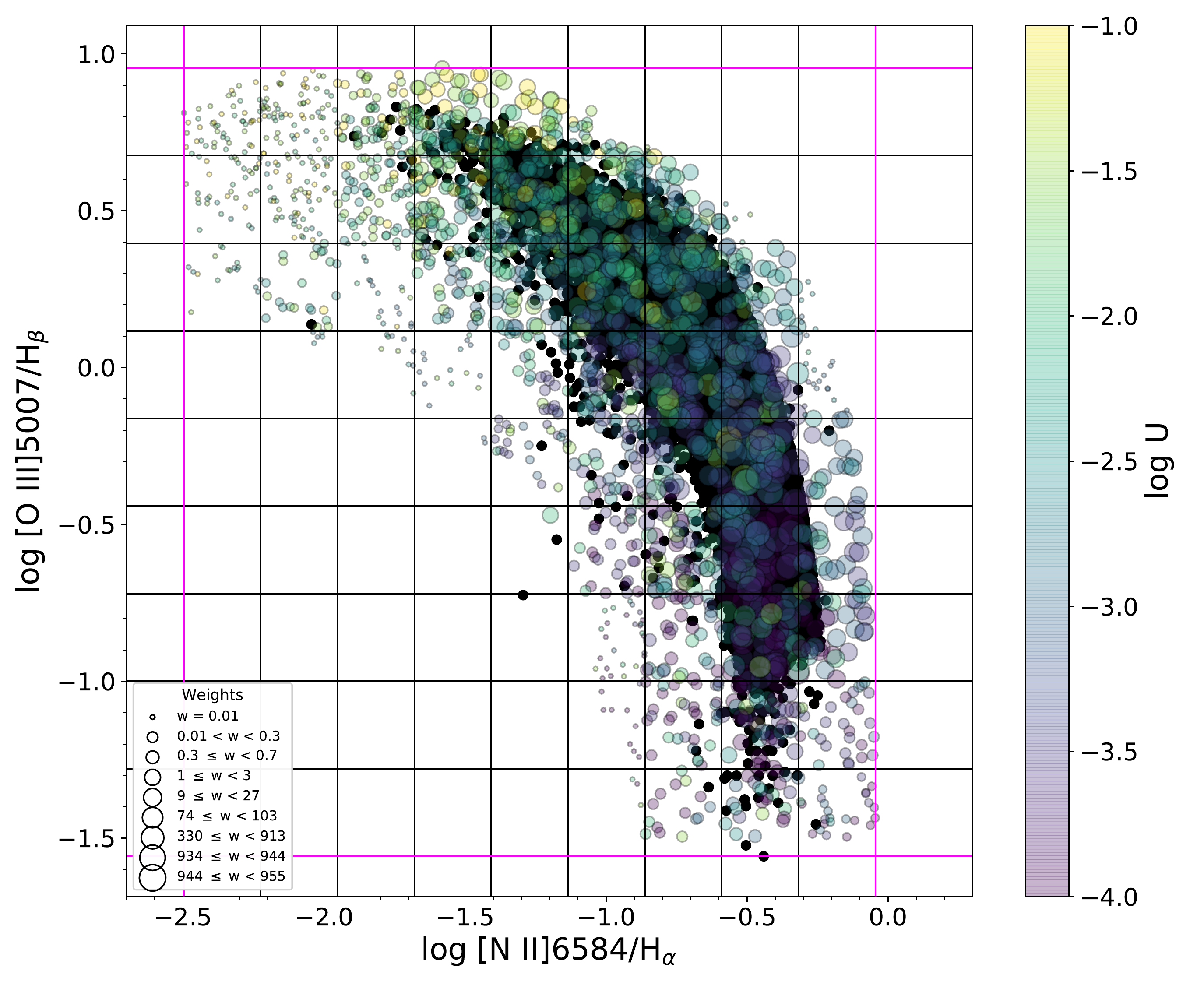}
\caption{Values of [\ion{O}{iii}] $\lambda$5007/H$\beta$ as a function of [\ion{N}{ii}] $\lambda$6584/H$\beta$ (the BPT diagram). The parameter space is divided in cells. The color dots represent the weighted models (the size of the point is related to the weight) whereas the black dots represent the observed extragalactic \ion{H}{ii} regions.} 
\label{fig3}
\end{center}
\end{figure}

Medina-Amayo et al. performed fits to obtain the best possible ICFs. One improvement with respect to the ICFs obtained by \citet{del2014} is that Medina-Amayo et al. associated a weight to each model. To do this, the parameter space of Fig.~\ref{fig2} was divided in cells and a weight was associated to each model according to the number of observations in each cell (see Fig.~\ref{fig3}). For example, a model located in a cell where there are no observations will have a very low weight whereas a model located in a cell where there are many observations will have a high weight. 

\begin{figure}
\begin{center}
\includegraphics[height=6.5cm]{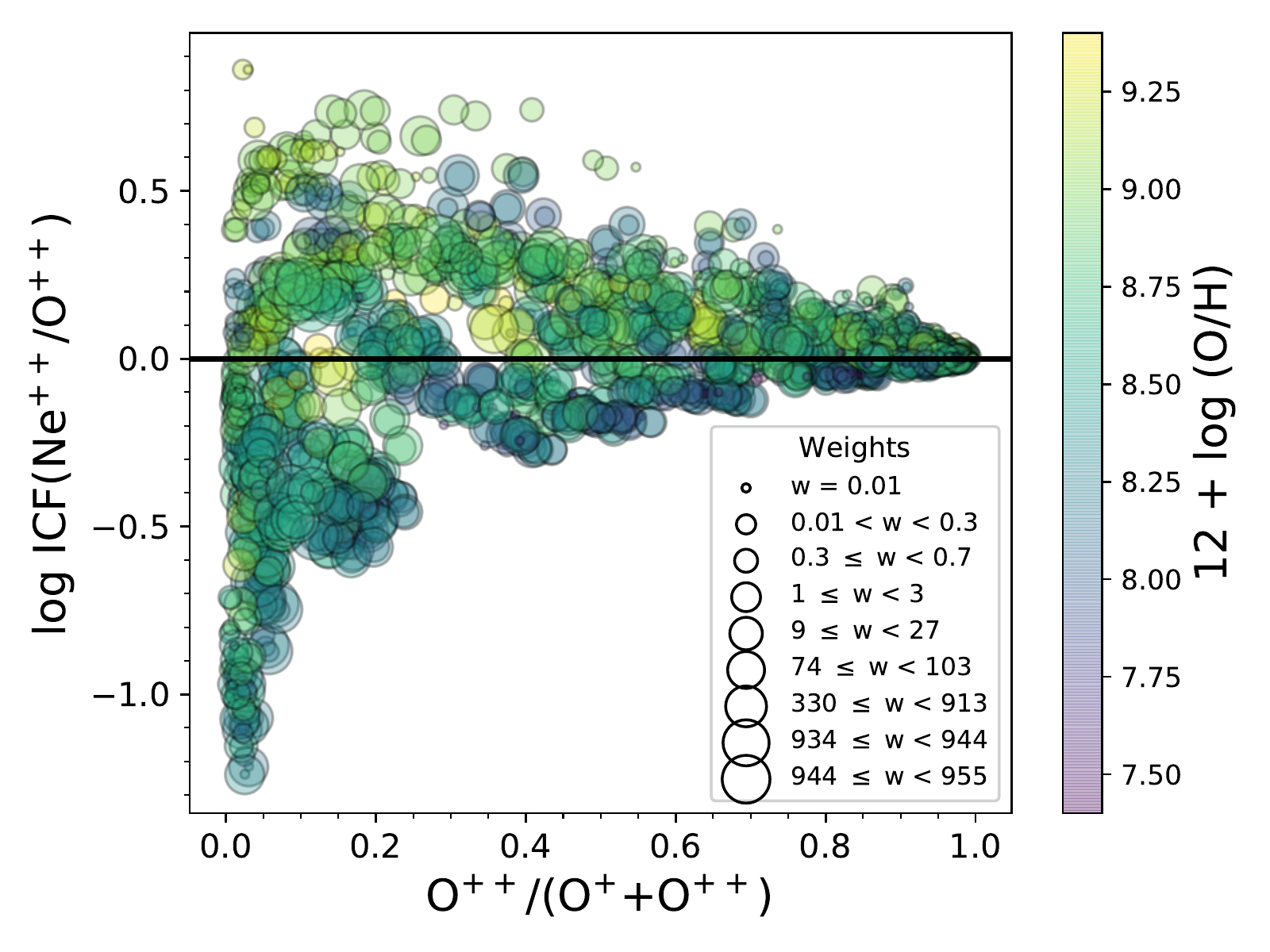}
\caption{Values of $x$(O$^{++}$)/$x$(Ne$^{++}$) as a function of O$^{++}$/(O$^{+}$+O$^{++}$) for our sample of weighted photoionization models representative of extragalactic \ion{H}{ii} regions. The colorbar runs from low lo high values of the O/H value. The black line represents where $x$(O$^{++}$)/$x$(Ne$^{++}$) = 1 as assumed by the classical expression Ne/O = Ne$^{++}$/O$^{++}$.} 
\label{fig4}
\end{center}
\end{figure}

The ICFs were computed by fitting analytical expressions to the models taking into account their weights. The use of the weights not only allows one to obtain a value of the ICFs more representative of the bulk of the observations but also to obtain more realistic expressions for their uncertainties. In principle, the weighted ICFs should be more adequate to obtain total abundances. Figure~\ref{fig4} shows the values of $x$(O$^{++}$)/$x$(Ne$^{++}$) as a function of O$^{++}$/(O$^{+}$+O$^{++}$) for the weighted models. 

The whole analysis and results for C, N, O, Ne, S, Cl, Ar will be presented by Medina-Amayo et al. (2019, in prep.).

\section{Discussion about ICFs}
\label{discussion}

\begin{figure}
\begin{center}
\includegraphics[height=5cm, trim = {-45 5 35 5}, clip]{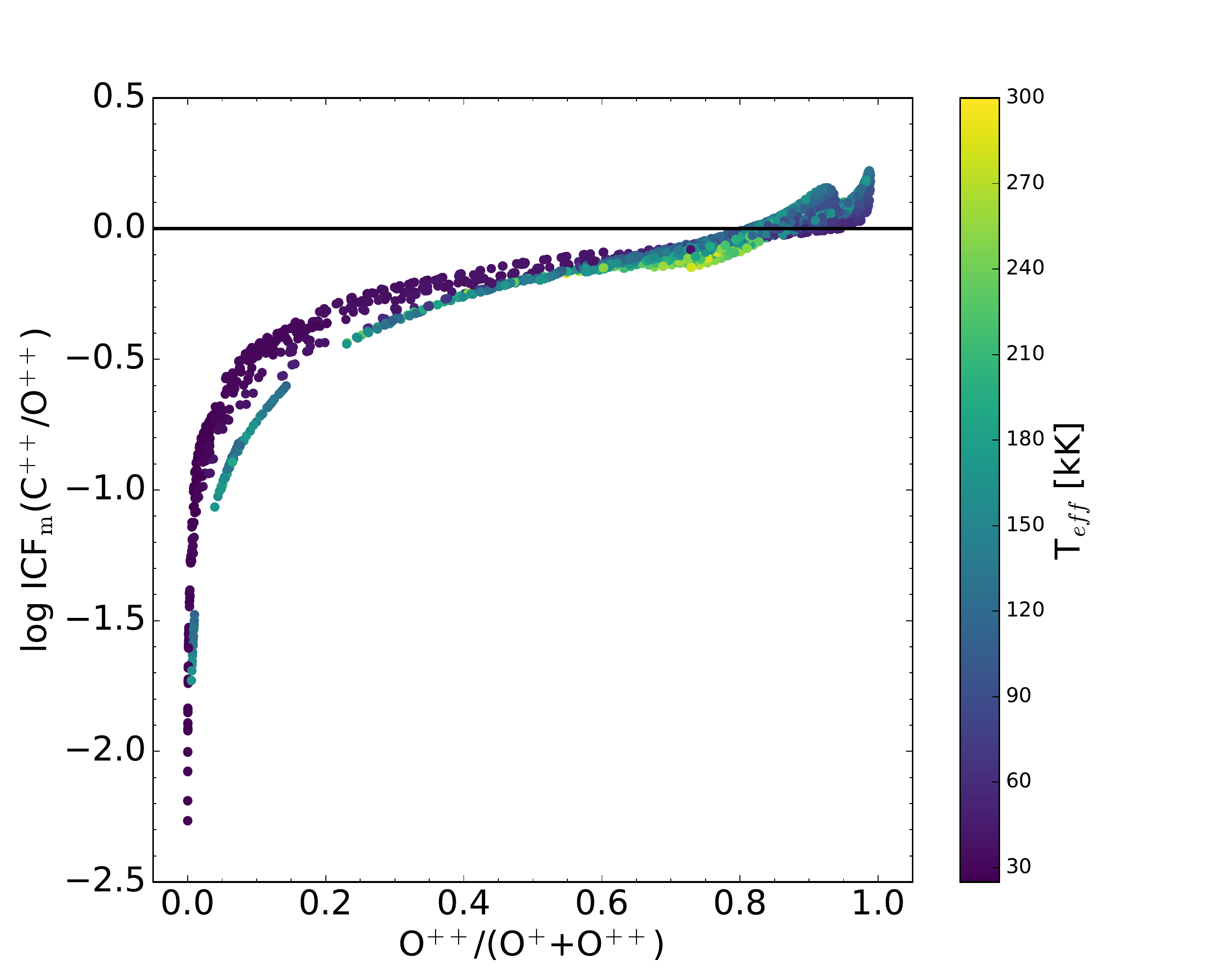}
\includegraphics[height=5cm, trim = {-45 5 35 5}, clip]{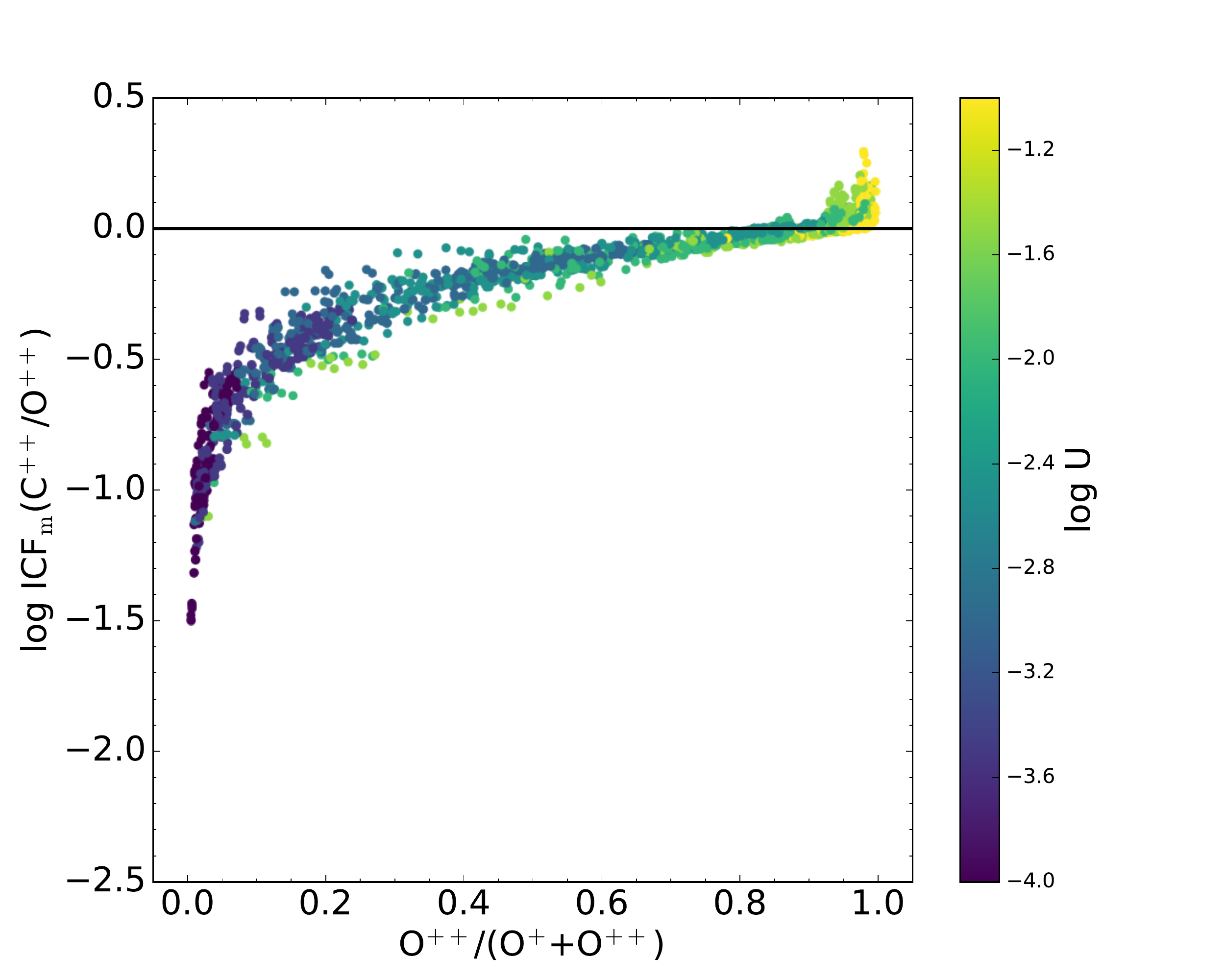}
\caption{Values of $x$(O$^{++}$)/$x$(C$^{++}$) as a function of O$^{++}$/(O$^{+}$+O$^{++}$) for our sample of photoionization models representative of PNe (left panel) and extragalactic \ion{H}{ii} regions (right panel). The colorbars run from low lo high values of the effective temperature of the central star (for the PN models) and of the ionization parameter (for the extragalactic \ion{H}{ii} regions).}
\label{fig5}
\end{center}
\end{figure}

A good ICF is one that does not introduce any bias or spurious trend in the derived abundances. One can use photoionization models to explore possible biases introduced by empirical ICFs based on ionization potential considerations. For example, from the grid of models presented in Sections~\ref{icfs_pne} and \ref{icfs_hii} we obtain that using the expression C/O = C$^{++}$/O$^{++}$ may overestimate C/O values in up to 2.3 and 1.5 dex in PNe and \ion{H}{ii} regions, respectively, in low ionization nebulae (see Fig.~\ref{fig5}). 

Sometimes there is no obvious trend between the total abundances and the degree of ionization but an unexpected trend appears while using the derived abundances. This is the case for N/O. The ICF proposed by \citet{del2014} does not introduce any trend between N/O and O$^{++}$/(O$^{+}$+O$^{++}$). However, \citet{del2015} found that this ICF seems to introduce an artificial trend between N/O and He/H values in PNe and thus, it is preferable not to use it.

\begin{figure}[!h]
\begin{center}
\includegraphics[height=9cm, trim = {0 80 0 80}, clip]{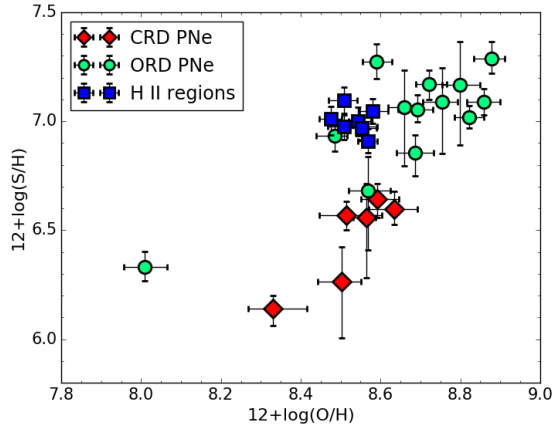}
\vspace{-1.7cm}
\caption{Values of S/H  as a function of O/H for the sample of PNe and \ion{H}{ii} regions studied by \citep{del2015}. The red diamonds represent PNe with C-rich dust, the green circles represent those with oxygen-rich dust, and the blue squares represent a group of Galactic \ion{H}{ii} regions. }
\label{fig6}
\end{center}
\end{figure}

The ICF may be wrongly invoked as responsible for unexplained results. For example, \citet{hen2004} suggested that the ICF is the most likely cause of the sulfur anomaly: the fact that, for the same O/H value, PNe systematically show lower S abundances than \ion{H}{ii} regions. However, Fig.~\ref{fig6} shows that using a different ICF (the one proposed by \citealt{del2014}) does not solve this problem so the explanation is likely different. This group of objects include Galactic PNe with C-rich dust,  Galactic PNe with oxygen-rich dust, and a group of Galactic  \ion{H}{ii} regions. 

In some cases one can compare abundances obtained using ICFs with those obtained by summing up all the ionic abundances of a given element. S. R. Pottasch and J. Bernard-Salas have done a lot of work in computing chemical abundances by directly adding individual ionic abundances (without using an ICF) of PNe from infrared, ultraviolet, and optical spectra \citep[see, e.g.,][]{pot2008, pot2009a, pot2009b, pot2011}. However, uncertainties associated with corrections for aperture effects may not be negligible.

\citet{rod2005} computed iron abundances in a group of PNe and \ion{H}{ii} regions by 1) adding up the ionic abundances of Fe$^{+}$, Fe$^{++}$, and Fe$^{+3}$ obtained from the emission lines of [\ion{Fe}{ii}] (in some cases), [\ion{Fe}{iii}], and [\ion{Fe}{iv}] and 2) by using Fe$^{+}$ and Fe$^{++}$ abundances and a theoretical ICF derived by them from photoionization models. They found a significant discrepancy between the empirical and theoretical ICF and concluded that the most likely explanation for this discrepancy is the inadequacy of some of the atomic data of iron. 

\begin{figure}[!h]
\begin{center}
\includegraphics[height=9cm, angle=270, trim = {0 0 0 10}, clip]{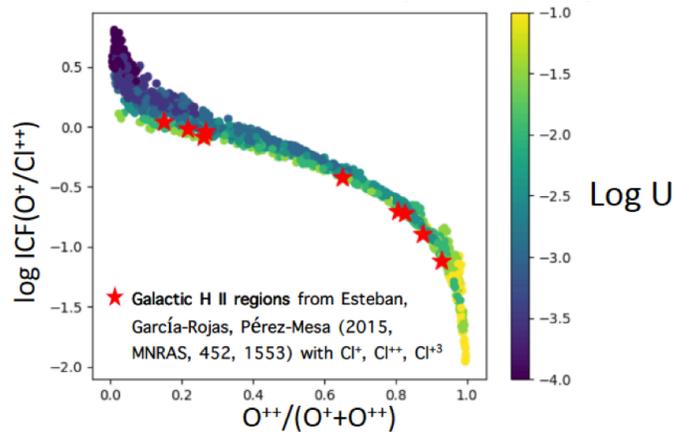}
\vspace{-0.9cm}
\caption{Values of $x$(O$^{+}$)/$x$(Cl$^{++}$) as a function of O$^{++}$/(O$^{+}$+O$^{++}$) for the sample of models representative of extragalactic \ion{H}{ii} regions (color dots). The red stars represent a group of Galactic H II regions observed by \citet{est2015} where Cl$^{+}$, Cl$^{++}$, and Cl$^{+3}$ can be computed to obtain the empirical ICF for chlorine. The colorbar runs from low lo high values of the ionization parameter.}
\label{fig7}
\end{center}
\end{figure}

\citet{est2015} computed chlorine abundances in a group of Galactic \ion{H}{ii} regions directly by adding up the abundances of Cl$^{+}$, Cl$^{++}$, and Cl$^{+3}$ and provide an empirical ICF. We illustrate in Figure~\ref{fig7} the values of $x$(O$^{+}$)/$x$(Cl$^{++}$) as a function of O$^{++}$/(O$^{+}$+O$^{++}$) for our sample of extragalactic \ion{H}{ii} region models (color dots) and also for the group of  \ion{H}{ii} regions studied by \citet{est2015} (red stars). It is clear that a fit to the models will lead to an ICF (and hence, a Cl/O value) somewhat higher than a fit to the observations. One possibility is that photoionization models are not completely adequate to describe ionization structure of Galactic \ion{H}{ii} regions but another possibility is that the ionic abundances computed from observations are not correct (for example, because of using an incorrect \temp, see papers by Rodr\'iguez and Dom\'inguez-Guzm\'an et al. in these proceedings). Therefore, still some work has to be done with models and observations to find the best ICF for each element.

\clearpage

\section{Conclusions}
We want to end by mentioning again that ICFs are essential to compute the abundances of many elements (the only alternative is to compute a detailed photoionization model). In principle, ICFs derived from photoionization models should be better than empirical ones based on ionization potential considerations because photoionization codes include all the relevant physics involved in ionized nebulae. However, there are a few considerations that have to be kept in mind. Theoretical ICFs rely on idealized photoionization models whose structure may be different from real objects, on atomic data which may be incomplete and sometimes incorrect, and on a description of the ionizing radiation field which relies on stellar atmosphere models and hypotheses regarding the distribution of stellar masses and ages in the case of giant \ion{H}{ii} regions. Observations may help to test and refine theoretical ICFs. However, the ionic abundances derived from observations have their own problems, as shown in several contributions to these proceedings. 

\acknowledgments G. D. I. wants to thank the organizers of this event for inviting her to give this talk and to all the participants for making this workshop very fruitful. The authors would like to thank support from PAPIIT (DGAPA-UNAM) grant no. IA-101517 and CONACyT grants 241732 and 254132.

\bibliographystyle{aaabib}
\bibliography{bib}

\end{document}